\documentclass[5p]{elsarticle}

\usepackage{lineno,hyperref}
\modulolinenumbers[5]

\usepackage{float}
\usepackage{graphicx}

\journal{Surface Science}

\usepackage{amssymb}

\begin{document}
\begin{frontmatter}
\title{An electron acceptor molecule in a nanomesh: F$_{4}$TCNQ on $h$-BN/Rh(111)}
\author[mymainaddress,mysecondaryaddress]{Huanyao Cun\corref{mycorrespondingauthor}}
\cortext[mycorrespondingauthor]{Corresponding author}
\ead{hycun1@physik.uzh.ch}
\author[mytertiaryaddress,mytertiaryaddressb]{Ari Paavo Seitsonen}
\author[mymainaddress,myquaternaryaddress]{Silvan Roth}
\author[myquinaryaddress]{Silvio Decurtins}
\author[myquinaryaddress]{Shi-Xia Liu}
\author[mymainaddress]{J\"urg Osterwalder}
\author[mymainaddress]{Thomas Greber}

\address[mymainaddress]{Physik-Institut, Universit\"{a}t Z\"{u}rich, Winterthurerstrasse 190, CH-8057 Z\"{u}rich, Switzerland}
\address[mysecondaryaddress]{Institute of Bioengineering, \'{E}cole Polytechnique F\'{e}d\'{e}rale de Lausanne, Route Cantonale, CH-1015 Lausanne, Switzerland}
\address[mytertiaryaddress]{D\'epartement de Chimie, \'Ecole Normale Sup\'erieure, 24 rue Lhomond, F-75005 Paris, France}
\address[mymainaddressb]{Institut f\"{u}r Chemie, Universit\"{a}t Z\"{u}rich, Winterthurerstrasse 190, CH-8057 Z\"{u}rich, Switzerland}
\address[myquaternaryaddress]{Institut de Physique, \'{E}cole Polytechnique F\'{e}d\'{e}rale de Lausanne, Route Cantonale, CH-1015 Lausanne, Switzerland}
\address[myquinaryaddress]{Department of Chemistry and Biochemistry, University of Bern, Hochschulstrasse 6, CH-3012 Bern, Switzerland}






\begin{abstract}

The adsorption of molecules on surfaces affects the surface dipole and thus changes in the work function may be expected. The effect in change of work function is particularly strong if charge between substrate and adsorbate is involved.
Here we report the deposition of a strong electron acceptor molecule, tetrafluorotetracyanoquinodimethane C$_{12}$F$_4$N$_4$ (F$_{4}$TCNQ) on a monolayer of hexagonal boron nitride nanomesh ($h$-BN on Rh(111)). The work function of the F$_{4}$TCNQ/$h$-BN/Rh system increases upon increasing molecular coverage. The magnitude of the effect indicates electron transfer from the substrate to the F$_{4}$TCNQ molecules. Density functional theory calculations confirm the work function shift and predict doubly charged F$_{4}$TCNQ$^{2-}$ in the nanomesh pores, where the $h$-BN is closest to the Rh substrate, and to have the largest binding energy there. The preferred adsorption in the pores is conjectured from a series of ultraviolet photoelectron spectroscopy data, where the $\sigma$ bands in the pores are first attenuated. Scanning tunneling microscopy measurements indicate that F$_{4}$TCNQ molecules on the nanomesh are mobile at room temperature, as "hopping" between neighboring pores is observed.

\end{abstract}

\end{frontmatter}

\section*{Introduction}

Electron transfer is an essential process that governs many elementary processes in physics and chemistry. For atomic or molecular adsorbates on metallic surfaces simple arguments predict on whether charge is transferred from the surface to the molecule or vice versa. 
The relevant quantities are the work function of the surface and on the other hand the ionisation potential or electron affinity of the atom or molecule: If the work function is larger than the ionisation potential, an electron is transferred to the metal substrate and the adsorbate charges positively. This effect is well known and e.g. used for surface ionisation of alkali metal atoms \cite{Alton1993}. If the electron affinity is larger than the work function, an electron is transferred to the adsorbate from metal substrate \cite{Kocic2015}. 
In the case of surface ionisation it was pointed out by Gurney \cite{Gurney1935} that the quantum mechanical broadening of the adsorbate orbitals lead to the situation where the charge transfer is not complete and thus the electron tunnels back and forth between adsorbate and substrate. 
Also, the distance dependent interaction between substrate and adsorbate leads to a variation of the ionisation potential and the electron affinity, and the charge transfer itself changes the work function of the surface locally.

The charge transfer due to polarisation may be inferred from the induced change in the surface dipole. If the adsorbate density is known, the induced dipole may be determined from the Helmholtz equation that 
relates a change of the work function $\Delta \Phi$ with the areal density $n$ or coverage of the dipoles induced by the adsorbed species:
\begin{equation}
\Delta \Phi=-e~\frac{p~n}{\epsilon_0}
\label{E1}
\end{equation}
where $e$ is the elementary charge, $p$ the induced dipole parallel to the surface normal, and $\epsilon_0$ the vacuum permittivity.
From Eq.~\ref{E1} we see that the work function decreases if charge is transferred to the surface and oppositely increases if charge is transferred to the molecule.

If the metal surface is covered with a single layer graphite (graphene, $g$) \cite{Nagashima1994, Shikin1998} or hexagonal boron nitride ($h$-BN) \cite{Nagashima1995}, the reactivity of the system changes. In particular, the charge transfer time across the layer increases, and non-adiabatic effects may be observed \cite{Muntwiler2005}. 

The high electron affinity of the tetrafluorotetracyanoquinodimethane C$_{12}$F$_4$N$_4$ (F$_{4}$TCNQ; see the chemical structure in Figure 1a) has made it to be a frequently employed electron acceptor molecule, with which anion formation on surfaces may be studied. Previously, Barja et al. reported self assembly of TCNQ (C$_{12}$H$_4$N$_4$)
and F$_{4}$TCNQ (C$_{12}$F$_4$N$_4$) on the graphene/iridium system \cite{Barja2010}. At low temperatures different adsorption geometries and electronic properties, depending on where the molecules were located in the $g$/Ir(111) unit cell \cite{Kumar2017}.

The electron affinity of F$_{4}$TCNQ is 5.2 eV \cite{Gao2001} and exceeds the work function of the $h$-BN/Rh(111) nanomesh \cite{Dil2008} by 1.05 eV. Therefore we expect a complete electron transfer onto the molecule, and the formation of an open shell electronic system. This would open perspectives for spintronics as it was put forward for the TCNQ/$g$/Ru(0001) where long range magnetic order was found \cite{Garnica2013}.

Here we report on F$_{4}$TCNQ on the $h$-BN nanomesh on Rh(111). We have investigated the system with photoemission spectroscopy and scanning tunnelling microscopy (STM) on the same preparations.
Density functional theory (DFT) calculations complement the picture and come up with an unexpected prediction of structures with high adsorption energy and transfer of more than one electron onto the molecule.

The substrate in this study is the $h$-BN nanomesh on Rh(111) \cite{Corso2004}. It features a corrugated single layer of boron nitride \cite{Berner2007}, where 13$\times$13 BN units form a coincidence lattice on 12$\times$12 Rh units with the nearest neighbour distances (Figure 1b). The 12$\times$12 unit cell is constituted by 2~nm pores, where the nitrogen atoms are close to on-top Rh configuration, and where nitrogen lone pair bonding occurs \cite{Paffett1990}. These pores are separated by "wire" regions where the interaction of the $h$-BN with the substrate is weak. 
This pore-wire structure imposes strong {\it{lateral}} electrical fields (dipole rings) that are due to variations of the local electrostatic potential and that lead to to the self assembly of molecules in the pores \cite{Dil2008}.
The electrostatic landscape is such that negative charge is attracted to the pores, and accordingly we expect F$_{4}$TCNQ to first occupy the pores.

\begin{figure}[ht]
\begin{center}
\includegraphics[width=\columnwidth]{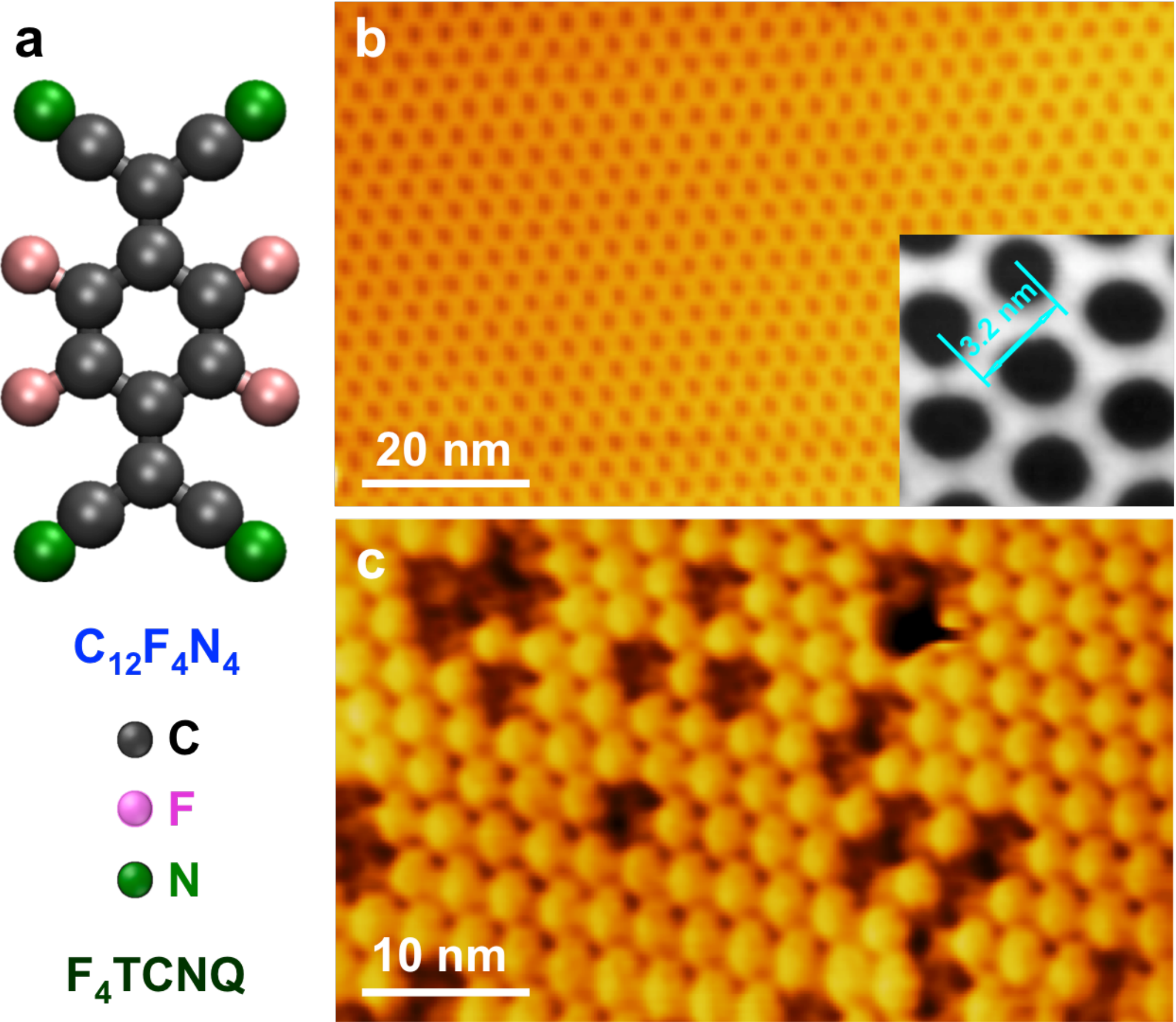}
\label{fig:Figure1}
\caption{{\textbf{\small F$_{4}$TCNQ on $h$-BN/Rh(111) surface.} (a) Ball and stick model of F$_{4}$TCNQ molecule (C$_{12}$F$_4$N$_4$). (b) and (c) Scanning tunnelling microscopy (STM) images at room temperature of pristine $h$-BN nanomesh on Rh(111). U$_t$ = -1.20 V, I$_t$ = 0.50~nA (b) and low coverage of F$_{4}$TCNQ molecules on $h$-BN/Rh(111) surface. U$_t$ = 1.0 V, I$_t$ = 0.3 nA (c). The right-bottom inset in (b) shows the supercell of $h$-BN nanomesh with a lattice constant of 3.2~nm. The wires appear bright and pores dark.
}}
\end{center}
\end{figure}

\section*{Results and Discussion}

Figure 1b displays STM images of the pristine $h$-BN nanomesh on Rh(111) substrate with a superstructure lattice constant of 3.2~nm (shown in the inset of Figure 1b). In Figure 1c an STM image of the $h$-BN/Rh(111) nanomesh after a small dose of F$_{4}$TCNQ with less than 1 molecule per nanomesh unit cell is shown, which corresponds to the data point marked with green square in Figure 2a.
"Bright" protrusions and "dark" depressions are imaged with the periodicity of the nanomesh. This imaging condition is obtained on a regular base after scanning the surface for some time at room temperature. The picture resembles the STM image of F$_{4}$TCNQ on $g$/Ir(111), where the protrusions have been assigned to the molecules \cite{Barja2010}. The ratio between dark and bright pores
and comparison to the coverage as determined via the X-ray photoemission spectroscopy (XPS) and the work function shift of 80 meV that were measured on the same sample rather favors a picture where the dark pores correspond to pores that are occupied with a molecule.

\begin{figure}[ht]
\begin{center}
\includegraphics[width=\columnwidth]{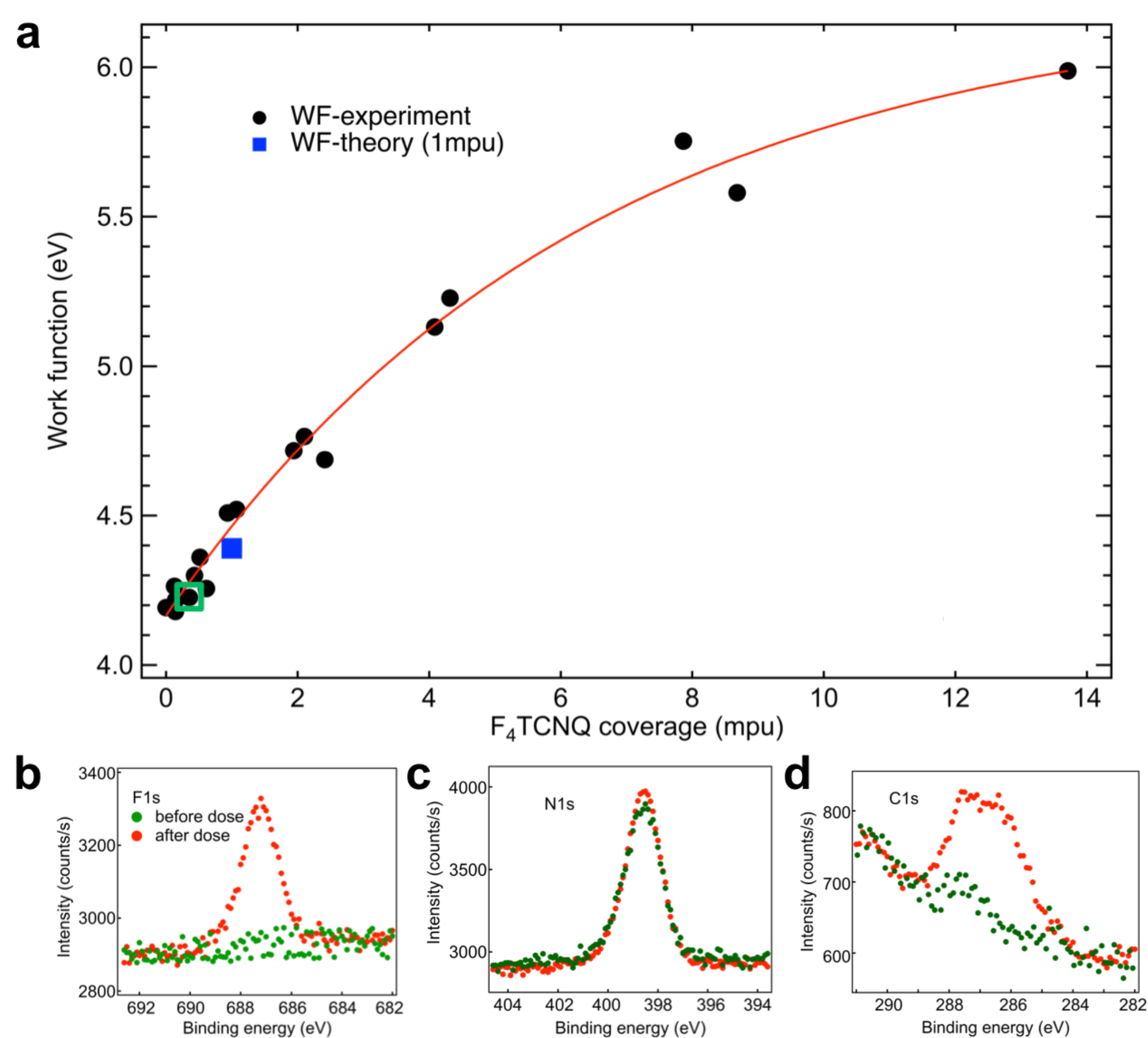}
\label{fig:Figure2}
\caption{{\textbf{\small Work function and XPS of F$_{4}$TCNQ/$h$-BN/Rh(111).} (a) Work function increase with increasing coverage of F$_{4}$TCNQ molecules. The black circles represent experimental data points at different F$_{4}$TCNQ molecular coverages, while the blue square indicates the calculated work function at F$_{4}$TCNQ coverage of 1 molecule per  unit cell. The green open square corresponds to the STM data shown in Figure 1c. The red line is the fit of an exponential function as described in the text. (b)-(d) Mg K$\alpha$ XPS ($\hbar\omega = 1253.6$ eV) spectra of F~1s (b), N~1s (c) and C~1s (d) before (green) and after (red) dosing F$_{4}$TCNQ on $h$-BN nanomesh (coverge of 7.8 mpu). After adsorbing on $h$-BN surface, F$_{4}$TCNQ shows distinct feature on F~1s and C~1s core levels.
}}
\end{center}
\end{figure}

Figure 2 shows the work function of the F$_{4}$TCNQ/$h$-BN/Rh(111) system as obtained from the width of He~I$\alpha$ excited normal emission ultraviolet photoemission spectra (UPS) as a function of molecular coverage $\Theta$. The coverage, given in units of molecules per nanomesh unit cell (mpu), has been determined from Mg~K$\alpha$ excited XPS intensity ratios between the F~1s, the N~1s and the C~1s peaks and the corresponding atomic photoemission cross sections, as shown at a coverage of 7.8 mpu in Figure 2b-2d. Accordingly we find a coverage of one F$_{4}$TCNQ mpu at an atomic F:N ratio of 4:173. Furthermore we note that the splitting in
the C 1s emission indicates the existence of chemically different carbon species in the molecule.

As expected for an electron transfer to the molecules the work function increases with coverage.
Fitting the data in Figure 2a to the function $\Phi=\Phi_0+\Delta\Phi(1-\exp(-\Theta/\Theta_0))$ we obtain the red line in Figure 2a and parameters $\Phi_0=4.2$~eV, $\Delta\Phi=2.07$~eV, and $\Theta_0=$~6.4~mpu.
From this we see that at coverages below 2~mpu the work function increases linearly.
The work function shift is much more pronounced than in the F$_{4}$TCNQ/$g$/SiC system where Chen et al. found an increase of 0.7~eV \cite{Chen2007}. 
Given the fact that the work function of $g$/SiC is similar to that of $h$-BN/Rh(111) this is an indication that the binding of F$_{4}$TCNQ is different to $h$-BN on a metal from $g$ on a semiconductor. 
For 1~mpu we obtain a work function shift of 0.3 eV at room temperature, and in cooling the sample to 185~K the work function further increases by 50~meV. The temperature dependent work function shift may be due to a higher occupancy in the pores.
These values can be compared to the result from the DFT calculations, where we obtain a value $\Delta\Phi_\mathrm{DFT,1\ mpu}$ of 0.19 eV as indicated with the blue square in Figure 2a. We consider this a very good agreement with experimental confirmation that at low coverage F$_{4}$TCNQ$^-$ anions are formed on the $h$-BN nanomesh.

Figure 3 shows normal emission He~I$\alpha$ UPS excited valence band photoemission data of F$_{4}$TCNQ on $h$-BN/Rh(111) as a function of the coverage $\Theta$. In Figure 3a a series of UPS spectra at different F$_{4}$TCNQ coverages from 0 to 4.3 mpu are shown. The spectrum at $\Theta$=0~mpu is recorded after annealing the clean nanomesh. 
Afterwards F$_{4}$TCNQ has been evaporated successively and measured with XPS and UPS after each evaporation. The spectra are dominated by the well-known valence band pattern of $h$-BN/Rh(111) with the Rh~4d valence band peak at 0.88 eV binding energy, and the split $h$-BN $\sigma$-band \cite{Berner2007}: $\sigma_\alpha$ that accounts for the electronic structure in the nanomesh wires and $\sigma_\beta$ for that in the nanomesh pores \cite{Goriachko2007}.

\begin{figure}[ht]
\begin{center}
\includegraphics[width=\columnwidth]{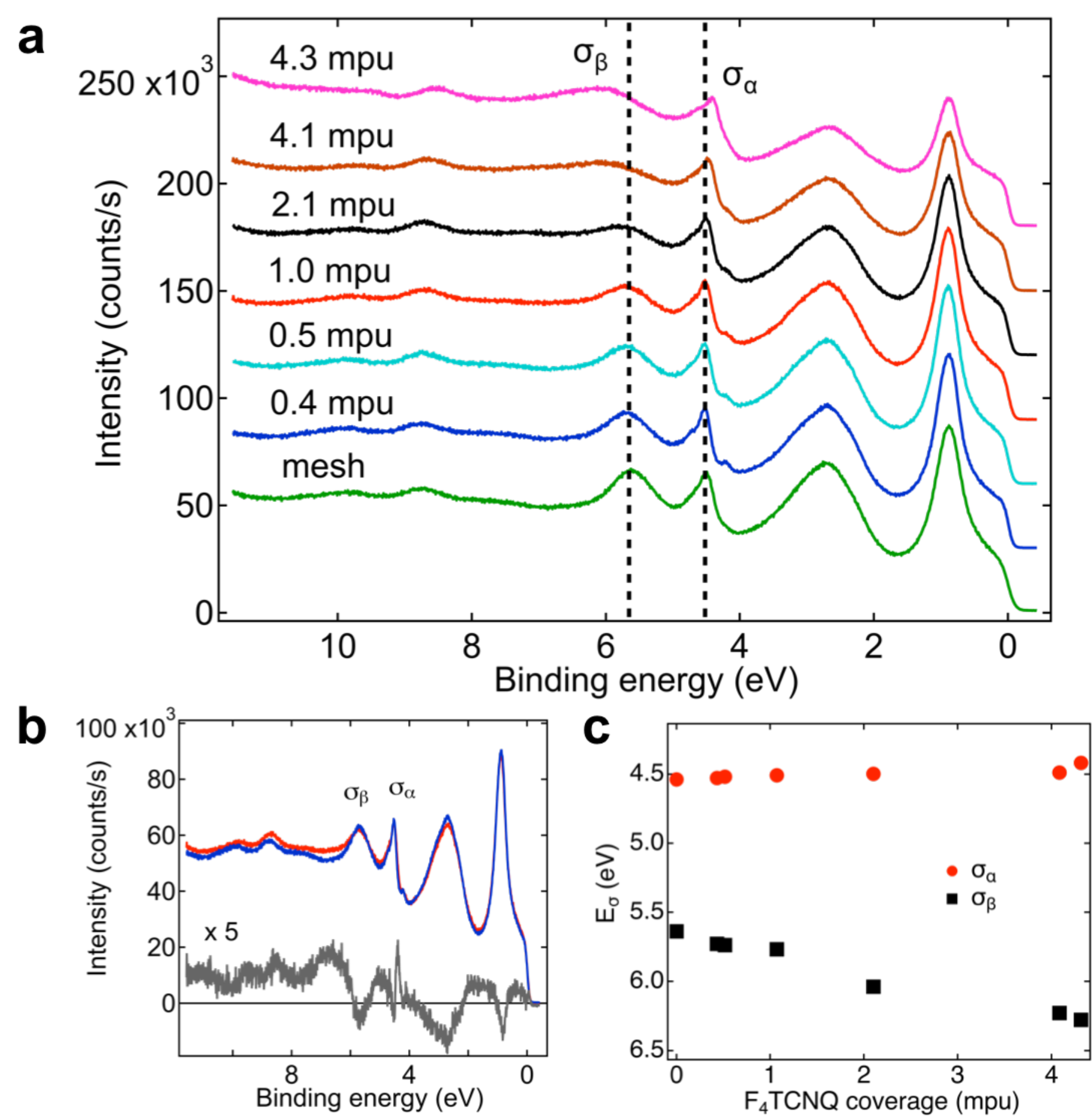}
\label{fig:Figure3}
\caption{{\textbf{\small He~I$\alpha$ normal emission ultraviolet photoemission spectrum (UPS) data of F$_{4}$TCNQ/$h$-BN/Rh(111).} (a) A series of UPS spectra of F$_{4}$TCNQ/$h$-BN/Rh(111) at different coverages of F$_{4}$TCNQ, showing that the $\sigma_\alpha$ and $\sigma_\beta$ bands shift upon increasing of F$_{4}$TCNQ molecules. The two black dashed lines are a guide to the eye. (b) Comparison of UPS spectra of 0.4 molecule per unit cell (mpu) (blue) and 1.0 mpu (red). The grey line is the difference between these two spectra (1.0 mpu-0.4 mpu). In order to better demonstrate the difference, the zero line is added, and the data are scaled by a factor of 5. Clearly, the $\sigma_\alpha$ and $\sigma_\beta$ bands are affected by adding more F$_{4}$TCNQ molecules, where the $\sigma_\beta$ band decreases significantly. (c) Shift of peak positions of $\sigma_\alpha$ and $\sigma_\beta$ bands with the increase of F$_{4}$TCNQ molecules. The $\sigma_\alpha$ band shifts towards lower binding energy, while the peak in the $\sigma_\beta$ band-region shifts to higher binding energy.
}}
\end{center}
\end{figure}

Like it was first shown for naphthalocyanine on $h$-BN nanomesh \cite{Berner2007} and later for Xe on $h$-BN nanomesh \cite{Dil2008}, adsorbates attenuate the photoemission signals from the substrate, and since the signal from the pore, $\sigma_\beta$ is attenuated before the signal from the wires $\sigma_\alpha$, a higher adsorption energy in the pores was concluded for naphthalocyanine and for 12 Xe atoms in the pores.
Here we observed the same trend: With increasing molecular coverage the $\sigma_\beta$ intensity decreases before it does the same in the the $\sigma_\alpha$ intensity. In Figure 3b the spectra of 0.4 mpu and 1.0 mpu are compared, and the difference between the two (1.0 mpu - 0.4 mpu) is displayed after scaling by a factor of 5. The above-said is confirmed, in particular the attenuation effect is also seen for the Rh substrate peaks.

The assignment of molecular orbitals is not possible at these low coverages, photon energy and emission angle.
We observe, however, a shift in binding energy for the boron nitride related $\sigma$ bands. The sharpest feature in the difference curve lies at the steep onset of the $\sigma_\alpha$ band at 4.39~eV binding energy and it is accompanied by a sharp negative peak at 4.53~eV. This indicates a $\sigma_\alpha$ band shift to lower binding energy with increasing coverage and is remarkable since the majority of the molecules must occupy the pores, as implied by the attenuation of the $\sigma_\beta$ intensity.
The spectral weight near the broader $\sigma_\beta$ peak appears to shift towards higher binding energies with increasing coverage. This may be related to the decrease $\sigma_\beta$ intensity and the increase of F$_{4}$TCNQ derived molecular orbitals at around 7~eV binding energy.
In Figure 3c the binding energy positions of the $\sigma_\alpha$ band and the local maximum near the $\sigma_\beta$ band are shown as a function of the F$_{4}$TCNQ coverage. For flat $h$-BN the sigma bands are known to align with the vacuum level and to lead to the "physisorption model" \cite{Nagashima1995}, where on the Rh(111) substrate the $\sigma_\alpha$ band has the same offset of about 9~eV from the vacuum level like on Ni, Pd and Pt. 
Therefore it is expected that with increasing work function the $\sigma_\alpha$ energy rises. The fact that the rise of 14 meV per mpu does not correspond to the 300 meV per mpu indicates again that the vast majority of the F$_{4}$TCNQ molecules are not adsorbed on the wires. The shift in $\sigma_\beta$ band of -161~meV per mpu may not be explained with the $h$-BN "physisorption model" of Nagashima et al. \cite{Nagashima1995}. It rather indicates that the F$_{4}$TCNQ$^-$ molecules influence the boron and the nitrogen atoms in the pore, and the above-mentioned photoemission intensity redistribution between $\sigma_\alpha$ and F$_{4}$TCNQ molecular orbitals that may impose as well an apparent shift.

Figure 4 shows STM images of a sample at low coverage measured with variable temperature (VT-)STM system. As mentioned above, low coverage of
F$_{4}$TCNQ/$h$-BN/Rh(111) displays "bright" and "dark" features on the $h$-BN surface at room temperature, which are confirmed by two different instruments, i.e., electron spectroscopy for chemical analysis (ESCA) Park Scientific (Figure 1c) and VT-STM (Figure 4).
Figures 4a-4f illustrate a sequence of continuous scanning of the same area with same scanning conditions, where the time intervals between the images is 125 s. The green and blue circles in each image indicate the same positions individually in order to guide the eyes. The 6 images evidently demonstrate F$_{4}$TCNQ "hopping" on the $h$-BN surface at room temperature. For instance, the "dark" depressions marked with green and blue circles in Figure 4d appear to be "bright" protrusions in the Figure 4e, while the same positions show the "dark" features again in the subsequent Figure 4f. In order to demonstrate more "hopping" events and in a longer time interval, we made a movie with 40 continuous STM images scanning with same condition on the same area in the Supporting Information. This "hopping" indicates high mobility of F$_{4}$TCNQ molecules on $h$-BN surface, which is similar to the case of F$_{4}$TCNQ on $g$/Ir, where the F$_{4}$TCNQ molecules were reported to "rotate" even at 77 K \cite{Barja2010}. 

\begin{figure}[ht]
\begin{center}
\includegraphics[width=\columnwidth]{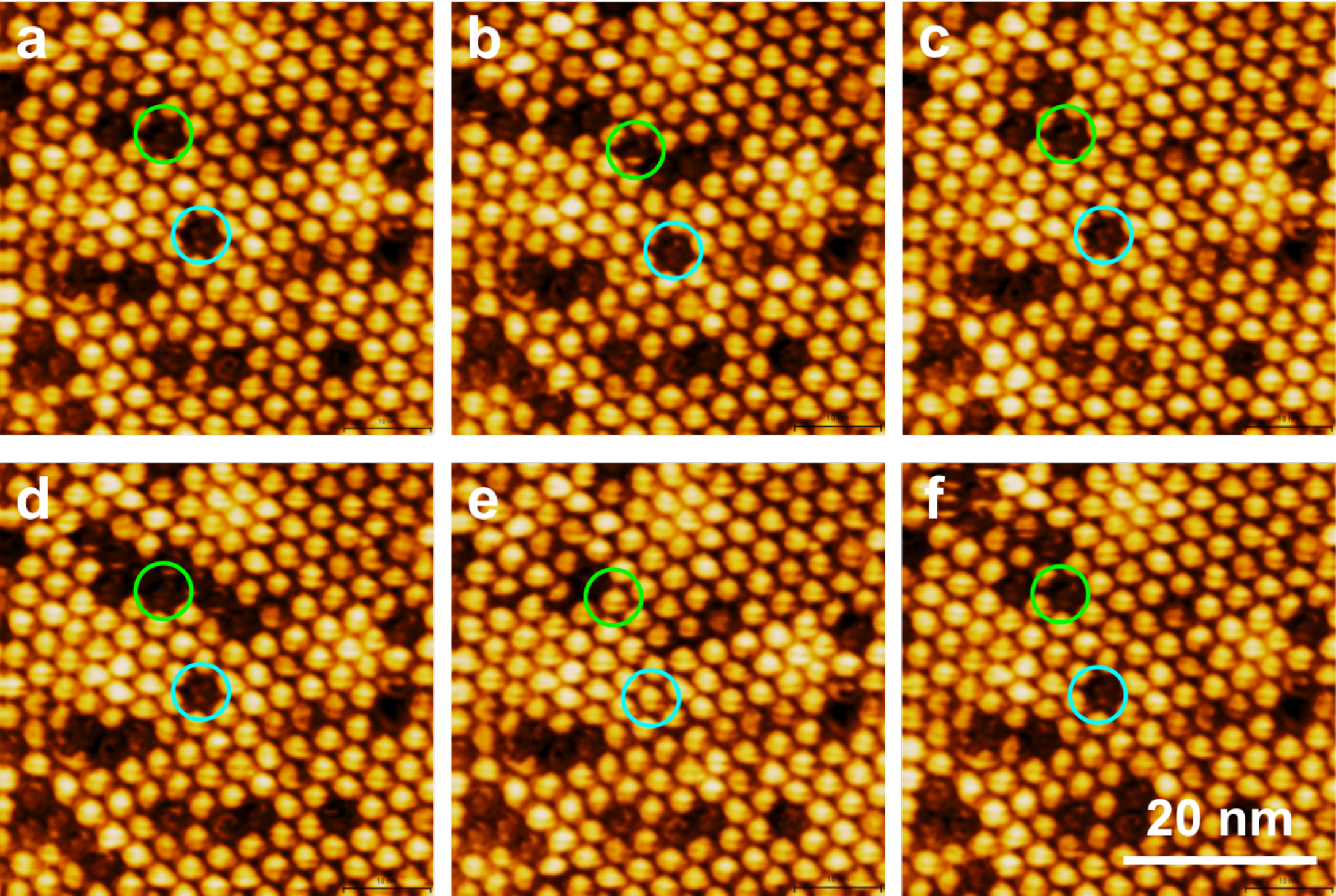}
\label{fig:Figure4}
\caption{{\textbf{\small F$_{4}$TCNQ "hopping" at low coverage below 1 molecule per unit cell.}
(a)-(f) A series of sequential STM images of F$_{4}$TCNQ on $h$-BN/Rh(111) at low coverage of the same area with the same scanning conditions show the F$_{4}$TCNQ molecules "hopping" (appear and disappear) on the surface at room temperature. The green and blue circles represent the same positions in every image to guide the eye. The time interval between two subsequent frames is 125 s. The contrast between dark and bright spots, as derived from the histogram of the $z$-values in Panel (d), is 2.6 \AA. U$_t$ = $-$1.0 V, I$_t$ = 0.005~nA.
}}
\end{center}
\end{figure}

To better understand the experimental results, we carried out DFT calculations. Because the molecular orbitals (for example the highest occupied molecular orbital, HOMO) preserve their character in the free molecule to a large extent, in the following we refer to those orbitals as in the neutral gas-phase molecule, even whn the initial lowest unoccupied molecular orbital (LUMO) gains electrons, to simplify the discussion. The DFT results confirm that the F$_{4}$TCNQ molecules like to occupy the pore of the $h$-BN nanomesh. The theory predicts the work function change and the charge transfer upon adsorption of one mpu. In Figure 5 the two lowest energy structures in the nanomesh pores are depicted. They are significantly distinct to each other, in binding energy, charge transfer, work function change and coordination to the beneath $h$-BN. In reference to the charge transfer the two structures are labeled as "$1-$" and "$2-$". 
Intriguingly, the work function change of "$2-$" is smaller than that of the "$1-$" structure, which must be related to the onset of covalent chemisorptive bonds and gain of exchange energy in the LUMO in "$2-$", which apparently superseeds the Coulomb repulsion in the LUMO. On the other hand "$1-$" is more physisorptive with ionic and van der Waals bond character and an open LUMO shell.
The work function change of 0.19~eV per mpu of the lowest energy "$2-$" structure fits well to the experimental result of 0.3~eV. 
Though, from the experiment we find no direct arguments on the charge state of F$_{4}$TCNQ on $h$-BN nanomesh. 
Rather, the high DFT binding energy difference between molecules in the pore and on the wire (see Table~1),  would not suggest much diffusion as we observe it at room temperature.
We have to speculate on how the detailed hopping mechanism as we observe it (see Figure~4) operates: Possibly the STM imaging process triggers hopping events.
In Figure 5c and 5f the orbital overlap $\mathcal{O}$, or the projection of the Kohn-Sham electronic states of the adsorption system on the molecular orbitals of a single molecule (but in the geometry as adsorbed) is shown for "$1-$" and "$2-$". The orbital overlap indicates at which energies the molecular orbitals reside when adsorbed, and if there are large changes, for example due to hybridisation. 
There are clearly defined states, in particular the HOMO$-$1, HOMO and LUMO, and the states above the LUMO are separated, like for the molecule in the gas phase. The gas phase eigenvalues are indicated with solid circles: Black for the occupied orbitals, and red for unoccupied ones. 
The energy alignment between gas phase and adsorbate system is arbitrary, and done to fit the HOMO and LUMO reasonably well. 
In the structure ``F$_4$TCNQ$^{2-}$'', where the Bader charge of the molecule is close to $-$2~$e$, the two fold degenerate, i.e. spin-degenerate LUMO of the gas-phase molecule is clearly below the Fermi energy, thus agreeing with the charge state "$2-$". Instead the charge state of ``F$_4$TCNQ$^{1-}$'' seems to pin at the Fermi level, indicating half filling. 

\begin{figure*}[ht]
\begin{center}
\includegraphics[scale=0.48]{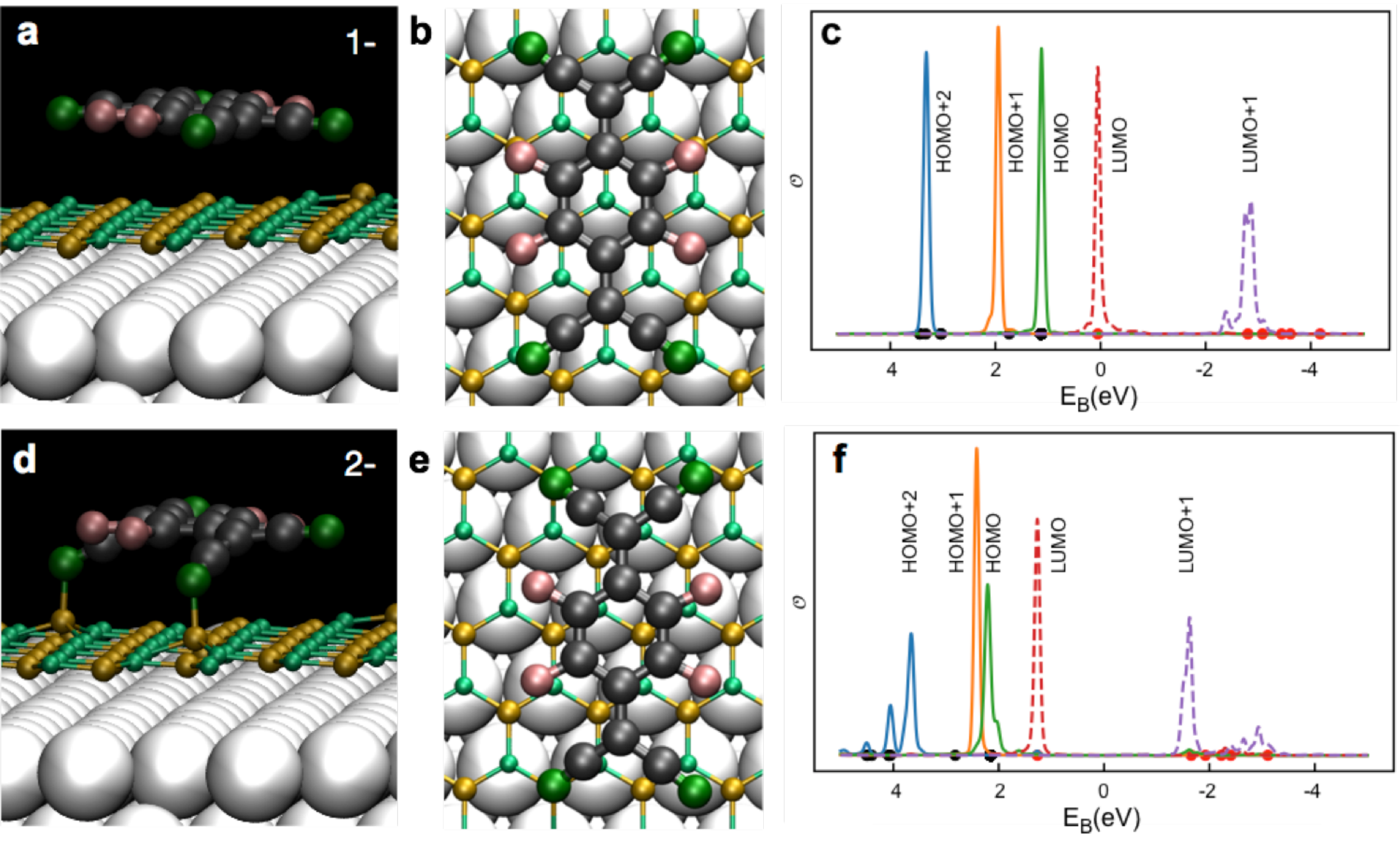}
\label{fig:Figure5}
\caption{{\textbf{\small DFT results of two different F$_{4}$TCNQ structures in the $h$-BN/Rh(111) nanomesh pores.} (a-c) The "$1-$" structure with one electron transfer. (a) Perspective view showing a floating molecule, (b) top view: The molecule is centered on top of the $h$-BN nitrogen atom with on top of Rh coordination and (c) the orbital overlap $\mathcal{O}$ with the LUMO pinned at the Fermi level. (d-f) The "$2-$" structure with two electrons transferred, that has a 26 \% higher binding energy than "$1-$". (d) Perspective view, showing the covalent bonding of two cyano (CN) groups to two boron atoms which get pulled off the $h$-BN layer. (e) Top view: The molecule is shifted away from the pore center such that two CN-B bonds become operational and (f) the orbital overlap $\mathcal{O}$, where the LUMO lies below the Fermi level. Note that the distinct distortion of the molecules imposes a non-rigid shift of the molecular orbital energies. The solid circles in (c) and (f) are the gas phase eigenvalues. Color code for atomic models: Boron orange, carbon black, fluorine pink, nitrogen green, rhodium whitish.
}}
\end{center}
\end{figure*}

\begin{table}[ht]
\caption{Calculated work functions $\Phi$ in 5 different structures. Binding energy $E_\mathrm{bind}$ in 4 different molecular conformations and the Bader charges on the molecules $q_\mathrm{Bader,molecule}$ in the 3 single molecule structures. \label{T1}}
\begin{center}
\begin{tabular}{l|ccccc}\hline\hline
System & $E_\mathrm{bind}$ (eV) & $\Phi$ (eV) & $q_\mathrm{Bader,molecule}$ ($e$) \\\hline
clean        & ---  & 4.20 & ---     \\
$1-$           & 3.07 & 4.61 & $-$1.34 \\
$2-$           & 4.15 & 4.39 & $-$1.89 \\
wire         & 1.78 & 4.88 & $-$0.40 \\
dimer (pore) & 3.64 & 4.88 
\\ \hline\hline
\end{tabular}
\end{center}
\end{table}

In Table~1 we also show the binding energy for a F$_4$TCNQ$^{2-}$ dimer. It is significantly lower than that of two molecules in two different pores. From this we expect single pore occupancy to be the lowest energy structure at low coverages.


\section*{Conclusions}

In conclusion, combining controlled experimental photoemission and scanning tunneling microscopy measurements with density functional theory calculations, we demonstrate electron transfer from $h$-BN/Rh(111) to F$_{4}$TCNQ. As negatively charged molecules (electron acceptors) coalesce on the surface, the work function of the system increases. Upon charging, the $\sigma_\alpha$ and $\sigma_\beta$ bands of $h$-BN nanomesh shift towards lower and higher binding energy, respectively. The molecules occupy preferentially the pores of the $h$-BN/Rh(111) nanomesh, though display mobility and molecular "hopping" on the $h$-BN surface at room temperature. DFT results indicate that the lowest energy structure is an anionic "$2-$" species. Our work paves the way to tune the electronic and structural properties of two-dimensional materials by using adsorption of organic molecules.

\section*{Methods}
\subsection*{Experimental}
The experiments were performed in two ultrahigh-vacuum (UHV) systems with base pressure of 1$\times$10$^{-10}$ mbar. One is a variable-temperature STM (Omicron VT-STM), and the other is a user-modified Vacuum Generators ESCALAB~220~ with a Mg K$\alpha$ lab source at an energy of $\hbar\omega = 1253.6$ eV,  with a monochromatized He~I$\alpha$ ($\hbar\omega = 21.2$ eV) source equipped with a room-temperature STM (Park Scientific) that allows photoemission and STM measurements on the same sample \cite{Greber1997,Auwarter2002}. The work function has been determined from the width of the He~I$\alpha$ photoemission spectrum recorded at a sample bias of -9~V.
The STM measurements were carried out with electrochemically etched tungsten tips. All STM images were taken in constant-current mode at room temperature. At negative tunneling voltages U$_t$ electrons are tunneling from the substrate to the tip.
The $h$-BN/Rh(111) samples were produced with the standard recipe \cite{Corso2004}. Particularly purified F$_{4}$TCNQ molecules were evaporated on $h$-BN/Rh(111) substrates kept at room temperature using a Knudsen cell at 365 K. Then the samples were transferred to the analysis chambers for the photoemission or the STM measurements.

\subsection*{Theory}

Calculations were performed using the
Kohn-Sham DFT formalism within the Gaussian plane wave (GPW) method \cite{Lippert1999} as implemented in the QuickStep module \cite{VandeVondele2005} in the CP2K program package \cite{CP2K}.
The exchange-correlation term was approximated with the rB86-vdW-DF2 exchange-correlation functional \cite{Hamada2014} that explicitly includes the van der Waals interactions.
In the GPW scheme we used the expansion of the molecular orbitals with Gaussian basis functions \cite{VandeVondele2007}, and the electron density is expressed with a plane wave basis set up to 700~Ry with a relative cut-off of 70~Ry.
Dual-space pseudopotentials~\cite{Goedecker1996} were used to describe the interaction of valence electrons with atomic cores.
We sampled the first Brillouin zone at $\Gamma$ point only. The theoretical lattice constant of Rh of 3.8147~\AA{} was employed in the calculations. The substrate was modelled with four layers of Rh(111), of which the two lowest were kept fixed at their bulk positions during the relaxation.

\section*{Conflict of Interest}
The authors declare no competing financial interest.

\section*{Acknowledgements}
Financial support by the Swiss National Science Foundation, by the EC under the Graphene Flagship (contract no. CNECT-ICT-604391) is gratefully acknowledged. We thank the Swiss National Supercomputer Centre (CSCS) for the generous allocation of computer time within the project \texttt{uzh11}.

\bibliography{F4TCNQ_BN-accepted-submit_to_arXiv}

\end{document}